# *Interface bonding of a ferromagnetic/semiconductor junction :*
# *a photoemission study of Fe/ZnSe(001)*

M. Eddrief, M. Marangolo and V. H. Etgens

Institut des NanoSciences de Paris, INSP-UMR CNRS 7588, *Université Pierre et Marie Curie -Paris 6* et Université Paris 7, Campus Boucicaut – 140 rue de Lourmel – 75015 Paris, France.

S. Ustaze and F. Sirotti

*Laboratoire pour l'Utilisation du Rayonnement Electromagnétique, CNRS-Université de Paris-Sud.*

M. Mulazzi*, G. Panaccione

*TASC laboratory INFM – CNR in Area Science Park, S.S. 14, Km 163.5, I-34012 Trieste, Italy*

D. H. Mosca

*Departamento de Física – UFPR, Centro Politécnico C. P. 19091, 81531-990 Curitiba PR, Brasil.*

B. Lepine and P. Schieffer

*Equipe de Physique des Surfaces et des Interfaces, UMR CNRS-Université n°6627, Bât. 11C, Campus de Beaulieu, F35042 Rennes Cédex, France*

*Abstract*

We have probed the interface of a ferromagnetic/semiconductor (FM/SC) heterojunction by a combined high resolution photoemission spectroscopy and x-ray photoelectron diffraction study. Fe/ZnSe(001) is considered as an example of a very low reactivity interface system and it expected to constitute large Tunnel Magnetoresistance devices. We focus on the interface atomic environment, on the microscopic processes of the interface formation and on the iron valence-band. We show that the Fe contact with ZnSe induces a chemical conversion of the ZnSe outermost atomic layers. The main driving force that induces this rearrangement is the requirement for a stable Fe-Se bonding at the interface and a Se monolayer that floats at the Fe growth front. The released Zn atoms are incorporated in substitution in the Fe lattice position. This formation process is independent of the ZnSe surface termination (Zn or Se).

The Fe valence-band evolution indicates that the *d*-states at the Fermi level show up even at submonolayer Fe coverage but that the Fe bulk character is only recovered above 10 monolayers. Indeed, the Fe $\Delta_1$-band states, theoretically predicted to dominate the tunneling conductance of Fe/ZnSe/Fe junctions, are strongly modified at the FM/SC interface.

**Introduction.**

Mixed ferromagnetic/semiconductor (FM/SC) heterostructures gain an increasing interest nowadays due to their potentialities for spintronic applications as spin injectors and spin analyzers for polarised currents into semiconductors.[1,2] The electronic and magnetic properties of these heterostructures depend on the nature of the metal/semiconductor interface and can affect magnetotransport.[3-5] In particular, the metal/semiconductors interfaces in FM/SC/FM heterostructures play the role of spin filters and determine the Tunnel Magnetoresistance (TMR) efficacy. The symmetry of the wave functions is the leading parameter for spin polarised transport : in the case of Fe electrodes, at the Fermi level, the iron majority $\Delta_1$ ($C_{4v}$) couple efficiently with the slowly decaying $\Delta_1$ ($C_{2v}$) state in the semiconductor. *In the theoritical treatments of tunneling, by assuming a perfect interface structure and two dimensional periodicity perpendicular to the direction of growth,* Fe/SC/Fe heterostructures can exhibit an extremely high degree of current spin polarization and a magnetoresistance approaching 100%.[3-5]

As a consequence, it is particularly relevant for spintronics applications of FM/SC heterostructures to determine the atomic and electronic structures at the interface. After the pioneer work of Jonker and collaborators[6], hybrid systems like Fe/GaAs and Fe/ZnSe have been extensively investigated.[7-10] Three important points were experimentally proven: (i) no reduction of the Fe magnetic moment (no magnetic dead layer) has been reported for the Fe/ZnSe interface.[8,9] This interface remains also stable against thermal annealing up to 300°C.[9] (ii) ZnSe requires a much lower substrate temperature for a good crystallization than GaAs.[11] This makes the tailoring of Fe/ZnSe/Fe heterostructures more comfortable.[12-14] (iii) Chemical reactivity is low (further discussed in this work) and the Schottky barrier is established after the growth of 2 ML of iron on ZnSe.[15]

These encouraging experimental findings hurt dramatically with spin-polarized tunneling magnetoresistance (TMR) values measured in Fe/ZnSe/Fe junctions : the measured values of tunneling magnetoresistance (TMR) were of about ~10% when polycrystalline ZnSe barriers were used[12] and even smaller, of the order of a few percent, in the case of a ZnSe crystalline epitaxial barrier.[13] TMR up to 16% was obtained by adding a few percent of Co to the Fe based layer.[14]

In this article we focus on the slight but measurable deviations of the Fe electrode and of the abrupt interface from the idealness. Our aim is double :

(i) to define the interface atomic environment where spin transmission takes place by core-levels X-ray Photoemission Spectroscopy (XPS) and Photoelectron diffraction (PED) (Section I). It turns out that even if the Fe/ZnSe junction presents an abrupt and stable interface, the ideal vision of a simple bulk-like Fe layer "softly landed" over the Zn-rich terminated ZnSe surface can be excluded.

(ii) To detect the electronic states relevant for TMR efficiency at the interface by valence band photoemission (Section II). We show that Fe-$\Delta_1$ ($C_{4v}$) state are appreciably modified by the interface. This is underestimated by previously reported *ab initio* calculations and it can determine the measured TMR values.

**Experimental details.**

This study has been carried out in three distinct experimental set-ups, two of them in synchrotron facilities. First, the core-levels and valence-band emission were measured with a high resolution spectrometer (Scienta 2002) located at the French synchrotron LURE (SB-7 SUPERACO beam line). The SUPERACO storage ring provided an intense photon flux that was selected in energy with a Dragon monochromator delivering photons with energy of 170 eV. The energy resolution of the whole set-up (incoming photons plus detector) was of about 0.15 eV, deduced from the measured width of the well-defined Fermi-level cut-off of a thick Fe film. The linear polarized photons impinge the sample surface with an incident angle of about 45° with respect to its normal. All spectra were measured by angle-integrated analyzer (± 6°) located at the surface normal emission position (Γ−Δ−H line).

Valence-band spectra were also collected at photon energy of 60 eV at APE-INFM surface laboratory and beam line at the ELETTRA storage ring of Sincrotrone Trieste (Italy). Sample preparation, energy resolution and measurement geometry were the same as at LURE. By using a photon energy of hν = 60 eV to minimize the photoelectron escape depth, optimum surface sensitivity is achieved. This photon energy also permits a direct comparison with previous work on valence-bands of bulk Fe(001) samples.[16]

X-ray photoelectron diffraction (PED) experiments were performed at PALMS Laboratory in University of Rennes (France) in a UHV chamber equipped by an Omicron EA125 hemispherical analyser with adjustable angular resolution. Mg K$_\alpha$ and Al K$_\alpha$ anodes (hν = 1253.6 and 1486.6 eV, respectively) were used to excite the Zn and Fe core-levels.

For all experiments, we have used an identical protocol with the iron deposition performed *in-situ* inside the corresponding UHV analysis chamber. ZnSe epilayers were first prepared in a molecular beam epitaxy (MBE) multi-chamber facility equipped with a III-V and a II-VI growth chambers interconnected. A 10 nm-thick of undoped ZnSe epilayer was grown on a GaAs buffer layer deposited on highly n-doped GaAs(001) substrate. At the end, the ZnSe epilayer was capped with an amorphous Se layer, a technique that has proved to be very efficient to prevent contamination during sample transportation.[9,15,17] Further details of MBE sample preparation can be found elsewhere.[11,15] Once in the photoemission set-up, samples were slowly heated up to 350°C in order to remove the Se capping layer and stabilize the c(2x2) Zn terminated surface. The iron was deposited using an e-beam evaporator at a growth rate of 0.3 Å/min. calibrated with a quartz microbalance and cross-checked by the core-levels photoemission decay. The growth temperature was kept constant at 180°C. The Fe coverage is expressed in monolayers (ML) with 1-(ML) $\equiv$ 1.21 x $10^{15}$ atoms/cm$^2$, corresponding to the atomic density of the bcc-Fe(001) plane, which is approximately twice the atomic density of the bulk terminated ZnSe (001) surface (and would produce a film ~ 1.4 Å thick if deposited uniformly).

We also performed experiments on Fe grown on a (2x1): Se ZnSe(001) surface. In the MBE chamber, we carefully stabilized the (2x1):Se-terminated surface, as attested by reflection high-energy diffraction (RHEED), by long annealing under Se flux in the MBE chamber. Since it is extremely difficult to stabilize this surface after de-capping amorphous Se epilayers, the uncapped 2x1 surface was transferred in a UHV portable chamber from MBE facilities to the photoemission beam line (SB-7, LURE). The surface cleanliness of transferred samples was checked by photoemission, showing the absence of carbon and oxygen traces. After the collection of these photoemission spectra, the symmetry of the (2x1) surface was further controlled by Low Energy Electron Diffraction(LEED).

### Growth and structure.

Firstly we recall previously reported details about the growth and the structure of Fe/ZnSe(001). Since the growth mode of Fe overlayers on ZnSe(001) is an essential information to interpret the photoemission results, we summarize rapidly some previous results. The survey of the RHEED diagram during the Fe growth shows that the initial C(2x2): ZnSe pattern, characteristic of the Zn-rich surface, rapidly fades into an increasing background with the persistence of the (1x1) pattern. Along the [110] direction, one observes the emergence of characteristic broad spots of Fe epilayer just after one monolayer thickness. Considering that the Fe lattice parameter is half of the ZnSe one, these Fe spots are superimposed with the ZnSe diagram on each two ZnSe streaks. Another feature of this Fe diagram is a reconstruction characterized by half order streaks along this direction. Considering an indexation of the Fe surface lattice with respect to the cubic bcc lattice, with $\mathbf{a}_{Fe}$ // [100] and $\mathbf{b}_{Fe}$ // [010], this reconstruction corresponds to a C(2x2) and it is known to be induced by the Se that is segregated at the Fe growth front.[18] At higher coverage, the Fe spots become elongated while keeping the C(2x2) Se-floating pattern (see Fig. 1 of ref. 15). This epitaxial relation has been confirmed by transmission electronic microscopy (TEM)[14] with (001) [110]Fe // (001) [110] ZnSe. The scanning tunnelling microscope (STM) results[17,19] have shown that at 1-ML coverage the film is composed of unconnected metallic clusters with an average diameter of ~ 30Å and a height of about 3Å. Extended x-ray-absorption fine-structure (EXAFS) measurements have shown that (i) the epitaxy-induced body-centred-tetragonal (bct) structure is recovered only at around 2 ML, with a very weak deformation with respect to the bcc bulk unit cell. (ii) The truncated semiconductor is continued by Fe layers, where half of the atoms in the (001) planes are positioned at crystal sites of the zinc-blende structure (bonding sites), and half of the atoms are located in the corresponding voids (antibonding sites).[17] Between 3 and 5 ML the islands lateral size increases leading to the formation of a continuous film. On the top, islands display an anisotropic shape, elongated along the [-110] direction.[17] At 7 ML the growth front roughness is strongly reduced.[17,19]

**Section 1: Interface atomic environment.**

In this section we will focus on the atomic environment close to the Fe/ZnSe interface by using X-ray Photoemission Spectroscopy (XPS) and X-ray Photoelectron diffraction (PED) techniques. By monitoring the core level components of Se *3d* and Zn *3d* as a function of Fe-coverage, XPS permit to detect the evolution of the interface, surface and bulk contributions (chemical shift and intensity attenuation). Complementary, PED will give usefull information about the atomic position of Zn atoms released on the Fe thin film. The experimental results reported below show that the ideal vision of a simple bulk-like Fe layer "softly landed" over the Zn-rich terminated ZnSe surface can be excluded. A simple consideration can be useful to anticipate our findings. The Fe- epilayers growth is obtained on a C(2x2):Zn-rich terminated ZnSe surface.[15,17] One Se monolayer is always found floating in the Fe growth front.[18] As a consequence, these boundary conditions concerning the interface formation demand atomic exchanges on two or three layers, at least.

**Core-levels.**

Figure 1 shows the evolution of the Se *3d*, Fe *3p* and Zn *3d* core-levels as function of Fe thickness for a deposition on a c(2x2):Zn reconstructed ZnSe(001) surface. Only selected spectra of representative Fe coverage are displayed. The spectra of the core-levels were first normalized to the photon flux. Next, in order to put into evidence the interface-induced effects, they have been aligned to their bulk-derived Zn and Se *3d* core-level components (within 50 meV) by a rigid shift to lower kinetic energy to compensate the band bending due to the Schottky-barrier formation.[15] A secondary electron background was also subtracted. After this standard procedure, core-levels were fitted using a standard non-linear least-squares procedure of symmetric Voigt line shapes for Se *3d* and Zn *3d* core-levels (a Daniach-Sunjic line-shape replaced the Lorentzian in the case of Zn-reacted species in metallic environment) and asymmetric line shape for Fe *3p*. The spin-orbit splitting was fixed to 0.85 and to 0.31 eV for Se *3d* and Zn *3d* spectra, respectively.[20] The branching ratio was fixed to 1.5. The remaining parameters, i.e., the peak positions, the relative intensity (height) and the full width at half maximum (FWHM), were allowed to vary freely.

The Fe-ZnSe interface formation process can be detected from the very first coverage (0.2 ML) since extra-components develop at the higher kinetic energy (lower binding energy) side of Zn *3d* and Se *3d* spectra, labeled as "$Zn^0$ released" and "Se-Fe", respectively. For higher Fe coverage (≥ 16 ML), a second Se component (labeled as "floating Se") becomes dominant. This component reflects the chemical environment of the Se floating layer on the metal growth front and a mild $Ar^+$ sputtering can completely eliminate this Se extra-component (see top spectrum in Fig. 1).

A line-shape decomposition of the spectrum (Fig. 1) provides a deeper understanding for the Fe/ZnSe interface formation process. The experimental peaks are described by ZnSe bulk and surface components (B and S, respectively) typical of the c(2x2):Zn-reconstructed ZnSe surface[20] plus the additional Fe-induced components ("Fe-Se", "floating Se" and "$Zn^0$ released"). No significant improvement was obtained by adding more components. Considering the metallic character of the Zn extra-component (Fig. 1), we have used an asymmetric line-shape. The origin of the gradual shift is not evident : an interpretation will be given after the presentation of PED data. Figure 1 shows the line-shape decomposition for Zn *3d* and Se *3d* core levels where good fits are obtained. However, we have observed a 20% Gaussian broadening of the Fe-induced components when compared with the ZnSe bulk peak. We attribute this effect to the disorder effects induced by the Fe/ZnSe interface formation.

The intensity evolution for each component of the Zn *3d* and Se *3d* peaks can be visualized in figure 2, determined after the integrated area and normalized to the total emission of the clean ZnSe surface. It is also plotted with dashed lines the behavior expected for an ideal layer-by-layer growth, without any interdiffusion or dissolution. We can notice that:

(i) The general behavior significantly runs off the ideal layer-by-layer growth curve;

(ii) The intensity of the ZnSe surface component diminishes rapidly due to the rapid disruption of the ZnSe reconstructed surface;

(iii) Up to 2 ML, the peaks corresponding to the ZnSe substrate are attenuated faster than the ideal case, in agreement with the proposed disruption of the topmost substrate layers;

(iv) The intensity of the Fe-induced components crosses a maximum at about 2ML and then decays exponentially, except for the "floating Se" component;

(v) The intensity of the "Se-Fe" component does not exceed ~ 30% of the total Se core-level intensity and after 8 ML-Fe coverage its signal asymptotically overlaps the substrate decay. The evolution of this component is compatible with one atomic Se terminating layer (the topmost ZnSe substrate) that bonds to the growing Fe epilayer;

(vi) The intensity of the Zn-reacted component ("$Zn^0$ released") grows up to ~ 2ML Fe nominal coverage and then decays more slowly than the substrate attenuation. This behavior agrees with released Zn atoms that are gradually incorporated into the Fe layer;

(vii) The exponential dependence of the Fe *3p* peak is consistent with the formation of a continuous and uniform ultra thin Fe film (see for example cross-section TEM image in Fig. 3 of Reference 14 and STM images of Reference 17).

Another interesting result was obtained for identical Fe growth procedure but performed over a (2x1):Se-reconstructed ZnSe surface. Surprisingly, no dependence of the substrate surface termination was detected. The thin film is still characterized by Se-Fe bond at the interface, Zn atoms diluted in Fe overlayer and one monolayer of Se atoms floating on the top surface of the metallic Fe film. These findings confirm that, irrespective of the initial ZnSe surface reconstruction, two atomic layers (Se-Zn) at the topmost of ZnSe substrate have been chemically converted during the Fe/ZnSe interface formation process.

**Photoelectron diffraction. Detection of Zn atoms released in the Fe thin film.**

If the phenomenological process for the Fe-ZnSe interface formation is quite clear in the light of photoemission results, it remains an open question concerning the released Zn atomic location inside the iron matrix. In order to get a deeper understanding on this process, we have employed the X-ray photoelectron diffraction (PED) technique for both Zn *3d* and Fe *3p* core-levels. It is fair to note that the mean free path for a classical photoemission experiment (XPS where the excitation source is hυ = 1253.6 eV with Mg anode) is three time larger than the adopted synchrotron source energy (hυ = 170 eV) and the experimental resolution is worse. As a consequence, the study of the Zn reacted component becomes much more difficult.

Nevertheless, we have succeeded to detect and to measure the "$Zn^0$ released" component by growing iron at room temperature. The room temperature growth probably quenches Zn atoms in off-equilibrium metallic configuration in the film leading to a less dispersed "$Zn^0$ released" peak than at 180°C. The " Zn° released" component is found at the lower binding energies side of the Zn *3d* substrate line (see insert in Fig. 3), shifted by ~1 eV relatively to substrate component. In order to determine the location of the Zn atoms expelled from the substrate after the Fe/ZnSe interface formation we have compared the Fe *3p* and Zn *3d* core levels PED polar scans along the [100] substrate direction. For the kinetic energies higher than 500 eV, due to forward scattering effects, strong intensity enhancements occur in the PED patterns along the denser atomic chains.[21] Here the kinetic energies for the Fe *3p* and Zn *3d* core levels are respectively 1.20 keV and 1.24 keV. For 12 ML Fe/ZnSe(001) we can see in Fig. 3 that the Fe 3p polar scan presents large peaks at 0 and 45°. These structures correspond respectively to forward scattering peaks along the [001] and [101] crystallographic direction of the Fe bcc lattice, in agreement with the observed α-Fe structure by EXAFS, even 2ML's iron.[17] The modulation of the Zn-reacted component show strong resemblances with that of the Fe *3p*. The same behavior is observed for 6 ML Fe deposited on ZnSe at RT. This result indicates that the Zn atoms liberated by the substrate are in the same crystallographic environment that the Fe atoms. .

The picture which emerges from these PED results is that the Fe/ZnSe interface forms a well-ordered Fe bcc phase with a solid solution of diffused Zn atoms gradually dissolved in substitution sites of the metallic matrix (see figure 4) (~0.5-1 ML Zn are dissolved in the 12 ML Fe film). Therefore, we can rule-out the formation intermetallic clusters like Zn-Fe in the Fe matrix (not observed in cross-section TEM image in Fig. 3 of Ref. 14). The anomalous broadening of the "$Zn^0$ released" contribution to the Zn *3d* spectra reported in the core-levels section can be interpreted as in the following: the substitutional Zn sites are not equivalent in this ultra thin bct Fe epilayer. Some Zn atoms are located at the interface, and feel a Se bonding in one side and Fe bonding in the other side. A few of them are located in the second layer and can still suffer of the interface influence, while others are embedded in the Fe film, far from the interface (figure 4). This creates local environments that are different but all metallic: As a consequence, Zn-reacted components are found in the low binding energy side (the "$Zn^0$ released" peak). In this context, the gradual energy shift of this reactivity-induced peak can be interpreted as the iron-induced onset of fully bcc metallic components, i.e. more and more zinc atoms in bcc sites, far from the interface.

**Discussion.**

A schematic picture of the Fe/ZnSe interface formation is reported in figure 4. This process can be summarized as follows: during the first iron deposition (around 2-ML), the two ZnSe topmost atomic layers (Zn and Se), plus the half Zn layer responsible for the Zn-rich c(2x2) ZnSe surface reconstruction[20], are disrupted and form a thin bct-Fe film with two most important characteristics: a stable Fe-Se bonding at the interface and a Se layer floating on the Fe film surface. The interface is then formed by a bulk truncated ZnSe terminated by Se that bonds to the Fe atoms. Fe atoms occupy crystal bonding and anti-bonding sites of the zinc-blende structure.[17] This Se-Fe bonding leads to the "Se-Fe" signature detected by photoemission that is shifted by about 0.8 eV towards the lower binding energies with respect to the bulk component. The chemical shift toward the lower binding energies is induced by the higher electronic density around Se atoms at interface as respect to the the ZnSe bulk. As predicted by Continenza et al.[22], charge flow tends to saturate the Se bonds by charge transfer from the Fe side into the ZnSe side. Moreover, the metallic character of the iron environment leads to a more effective core-hole screening. On the other hand, the binding energy of the "floating Se" component is very close to that of the ZnSe component. The distinct energy position observed for these two additional Se components reflects the difference of the charge readjustment of the Se atoms in two different environments: the first Se located at the interface is bonded to Zn down below and to Fe, up above. The second Se, located over the growth front of a metallic Fe layer (figure 3), is weakly bonded to the Fe overlayer. At larger coverage ($\geq$ 16 ML), Se *3d* core levels are still detected, even if the nominal Fe coverage is by far larger than the penetration depth. A short (5 min.) and mild sputtering succeed to remove a few topmost layers of the 28 ML-thick Fe film and results in a complete suppression of the Se signal, remaining only the signature of the metallic Fe *3p*. A detailed analysis helped us to assign an equivalent thickness for this segregating Se of approximately 1-ML of Se (in terms of ZnSe matrix). This sputtering experiment confirms that the remaining Se signal comes from the surface Fe film but also, within the detection limits of photoemission ($\approx$ 0.1%), attests also that selenium is not being incorporated inside the Fe overlayer.

A comparison with recent fully relaxed ab initio calculations is fruitfull. These calculations were performed for Fe/GaAs[23] and for Fe/ZnSe(001)[24]. It was shown that the interface of the Fe layer/semiconductor is abrupt and that at very low coverage the iron-anion bonding is energetically favored with iron in a substitutional sites. The following iron layers are found in the bcc stacking configuration. Furthermore, by an atomic exchange mechanism (not thermally activated), energy is minimized if an anionic layer floats on the Fe surface film (surfactant effect [18]) and interface iron atoms, bond to anions, occupy a substitutional site. Cations are predicted to be diluted in the Fe thin film. Our experiments confirm this theoretical scenario.

**Section 2. Electronic properties at the interface.**

The atomic environment of the interface has been obtained by photoemission core-levels measurements. In this section we present the valence bands evolution as a function of iron coverage. These measurements permit to detect interface electronic states that play an important role in TMR efficiency.

**Valence-bands .**

In figures 5 (a),(b) we present the valence-band evolution as function of iron coverage measured with two selected photon energies, 60 eV at ELETTRA and 170 eV at LURE. The raw data at submonolayer coverage are slightly shifted in order to take into account a residual ZnSe photovoltage effect. [15]

We can notice that at the very first iron deposition, the density of states (DOS) near the Fermi level ($E_F$) is entirely dominated by the Fe *3d* bands since the ZnSe contributions (*4s, 4p*) are located at higher binding energies, between 1.9 and 7 eV below $E_F$. The Fe *3d* contribution is enhanced due to its photoionization cross section that is two order of magnitude larger than those of ZnSe(*4s, 4p*) and Fe(4s, 4p).[25] As a consequence, the overall signal is very sensitive to Fe *3d* states from the very beginning of the Fe growth. We can clearly see this effect in the 0.2 ML-coverage spectrum (the lowest examined coverage) where the metallic character of the iron thin film is already visible. This very early metallic behavior is in agreement with the scanning tunneling spectroscopy (STS/STM) measurements done for coverage of 1ML-Fe.[17] We observe also that the ZnSe valence-band emission is rapidly attenuated as a function of the Fe thickness, coherently with core-levels photoemission reported above.

Close to the Fermi level, the valence band contribution is characterized by two overlapping structures: the first one is located just below $E_F$ (0.3 eV) and the second one at 0.8 eV (noted $P_0$ and $P_1$, respectively in fig. 5 (a),(b)). An additional peak labeled $P_2$ and located at 2.6 eV, is also clearly detected above 2-ML. We notice that the overall shape of spectra is strongly modified at low coverage and that the Fe-bulk valence-band is recovered only between 8 and 16 ML. In particular, the $P_1$ peak intensity development is frustrated at low coverage if compared with $P_0$ and $P_2$. Up to 4 ML-Fe coverage the relative intensity of these three peaks is roughly constant. We point also that *d*-band peaks are wider for the very low initial coverage and that they become narrower and more structured approaching the thickness at which the bulk iron character shows up. However, their energy position remains practically unchanged except for submonolayer iron coverage where the residual ZnSe photovoltage and band bending induce a small low-energy shift.[15]

In order to better visualize the electronic structure evolution as a function of the Fe thickness, we have plotted in figure 5 (c) the intensity ratio for $P_1/P_0$ and $P_1/P_2$ as a function of the Fe coverage, after subtracting the clean ZnSe contribution and without considering the important background above 3 eV. We observe that these ratios quickly evolve to the bulk values approaching 10 ML. This evaluation can be only qualitative because of the arbitrary shift of the ZnSe valence-band (residual photovoltage). A sputtering treatment (not shown) of a thicker Fe epilayer (28 ML) to remove the floating Se did not modify the valence band spectra. This corroborates the lack of surface states contribution in our spectra close to the Fermi level, as reported by other authors.[26] Indeed, surfaces states are expected to lie 0.2 eV and 2 eV above and below the Fermi level, respectively.

**Discussion.**

First, we summarize previous photoemission spectroscopy studies[16, 27] performed on the Fe bulk valence-bands. It is well known that the band spectral weight is strongly dependent on the angular integration, on the excitation energy and on the photon incidence angle with respect to the surface normal. The *d-band* features derived from relatively flat bulk bands near $\Gamma$ in the $\Gamma$-$\Delta$-H direction of the bulk Brillouin zone (normal emission) are constituted of two types of signatures: the first has a single band structure at about 0.3 eV below $E_F$ and the second display two structures located at 0.7 and 2.3 eV. Previous studies have assigned these peaks to minority and majority spin Fe *d*-bands states, $\Delta_{5down}$ at 0.3 eV, $\Delta_{1up}$ at 0.7 eV and $\Delta_{5up}$ at 2.3 eV, respectively. It is straightforward to correlate these signatures with the peaks $P_0$, $P_1$ et $P_2$ that have been discussed in the previous section. At normal emission, selection rules only allow initial states of $\Delta_1$ and $\Delta_5$ symmetry; $\Delta_1$ when the electric vector lies perpendicular to the surface, $\Delta_5$ when it is parallel. At high incidence angle (> 70°), the $\Delta_1$ majority peak (~ 0.7 eV) is relatively intense compared to the $\Delta_5$ exchange split bands (0.3 and 2.3 eV). On the contrary, the $\Delta_5$ peaks were expected to be stronger at small angles of incidence.[27] For the intermediary incidence angle, the valence-bands spectra are still dominated by the $\Delta_1$-peak at around 0.7 eV for both photon energies used in our experiments (60 eV [16], 170 eV [28]). Moreover, 60 eV photons permit to monitor the well-defined non dispersive *d*-bands near the Γ point.[16]

At 45° incidence angle, the valence-bands spectra of the thickest Fe film (28 ML) at the two photon energy (60 and 170 eV in Fig. 3) closely look like Fe bulk spectra[16,28] except for small changes in relative peak intensity and background.

The *first* consideration that we can do is that a simple visual inspection of the spectra indicates that no dramatic reduction of $\Delta_5$ band exchange splitting occurs for film thickness between 2 and 28 ML. This is in full agreement with the XMCD experiments[9] that have shown bulk-like magnetic properties for films thicker than 2 ML. After the subtraction of the ZnSe substrate contribution above 2 eV we note that the $\Delta_5$ band exchange splitting is roughly constant. No extra-peaks related to the reduction of the exchange splitting were observed. It is worthwhile to recall that a recent spin-resolved inverse photoemission experiment indicated the reduction of the spin-splitting of the Fe $\Delta_5$ bands near $H_{25}'$ for iron thickness below 30 ML.[19] The exchange splitting in Fe films is claimed to decrease from the expected 1.5 eV value for bulk iron to 1.2 eV at 15 ML, and to less than 1.1eV for a thickness of 8 ML. It is not the first time that contradiction between photoemission and inverse photoemission has been encountered. A similar case has been reported for the Fe/Ag(100) system, where the large reduction of the exchange splitting in Fe films was pointed by inverse photoemission spectroscopy [29] and not confirmed by photoemission spectroscopy for a 6 ML thick film.[30] It is a common thought that empty states are more sensitive to defects of the crystalline structure than occupied states. This assumption remains however with no conclusive experimental or theoretical evidence to our knowledge.

The *second* consideration, and maybe the most striking result of this photoemission valence-bands study, is the depression of the majority $\Delta_1$-states below 10 ML.
As shown above, core-levels photoemission results indicated that at 2 ML a stable Fe/ZnSe interface is established since no change of the line-shape spectra of Se and Zn 3d core-levels was detected (see Fig. 1 and 2). The overall picture that stands out consists of (i) a buried Fe-Se interface, of (ii) 1.5 ML of Zn atoms lost in the bcc Fe film and (iii) one Se layer floating on the growth front (see figure 4). As a consequence, the strong reduction of the $\Delta_{1up}$ - derived P1 peak intensity can be attributed to the electronic influence of the buried Fe/ZnSe interface. This interfacial electronic influence washes out beyond 10 ML and the bulk-like character is retrieved.

This unexpected $P_1$ reduction can be understood from the spatial orientation and extent of the $\Delta_1$ states. The $\Delta_1$-state have *s*, $p_z$ and $d_z^2$ wave functions symmetry. The $d_z^2$ and $p_z$ orbitals point towards the semiconductor so that a large electronic interaction with the substrate through $\Delta_1$-states is possible. As a consequence, interface Fe-Se hybridization leads to a depression of the Fe $\Delta_1$-state throughout the first Fe monolayers. On the other hand, the $\Delta_5$-states are characterized by in-plane extended $p_{x,y}$ and $d_{xz,yz}$ orbitals and weakly interact with the ZnSe substrate. We rule out any major influence of zinc atoms released in the Fe film on $\Delta_1$-state, for reasons linked to symmetry and energy levels arguments: Zn *3d* and Zn *4s* cannot strongly modify the Fe states close to the Fermi levels.

Indeed, *ab initio* calculations performed by different groups point that modifications of the local DOS are expected close to the FM/SC interface, mainly due to the dipole layer created by electrons transferred from Fe to the ZnSe.[3,4,31,32] This seems independent of the anionic[24,31] or cationic termination[4,32] of the semiconductor. Surprisingly, also the nature of the semiconductor seems not so important once that the same theoretical conclusions could be given for Fe/GaAs , Fe/ZnSe  and Fe/Ge interfaces. A major interface modification is that majority $\Delta_1$-states are depressed in the iron lead, close to the interface.[4,31,32] The extent of these interface induced modifications can take few layers.

By associating these *ab initio* calculations with our results we conclude that even if the trend of the strong depression of $\Delta_1$-states is described by calculation, its spatial extent seems experimentally wider than expected. Indeed, calculations localize this effect at the very first Fe layer, the second being much closer to bulk Fe. Only a larger extent of this effect, up to around 3-4 ML, can explain the strong $\Delta_1$ band depression observed at 8 ML. This could be possible if the screening of interface were less efficient than expected from the calculations. In this sense, we recall that calculations performed by E. de Jonge *et al.*[32] gave significant deviations of the local density of states up to three layers from the Fe/ZnSe interface. As a general result, we state that the interface electronic structure presents features that deviate from the idealness.

**Conclusion.**

We have studied the atomic environment of Fe/ZnSe interface and electronic properties of ultra thin Fe films, a well known "quasi ideal" metal/SC junction invoked as a good candidate for spintronics applications. By combining X-ray photoemission and photoelectron diffraction it was possible to draw a clear picture of the interface formation. The Fe bonding process starts with a disruption of two atomic ZnSe layers. In fact, two major statements must to be fulfilled: the Fe bonds to a Se terminated surface and one monolayer of Se floats in the growth front, independently of the ZnSe initial surface termination. The bcc Fe film accommodates around 1.5 ML of Zn atoms in bcc substitutional sites. These rearrangements are still very weak when compared with other FM/SC interfaces.

Close to the interface, Fe-electronic states are found strongly modified by the Fe-Se hybridization. In particular, the $\Delta_1$-band, because of the *s, $p_z$* and *$d_{z2}$* admixtures character, points towards the semiconductor and is strongly depressed if compared with the other *d*-bands features. We find that this band is more affected by the interface Fe-Se bonding than expected by calculations, since at 8 ML thick Fe film this effect is still clearly detected. The role of the interface bonding can be very important. Recent calculations have shown that in the case of Fe/MgO tunnel barrier, the presence of a Fe-O bonding can reduce the coupling between $\Delta_1$ states and the barrier and consequently the TMR.[33] The low TMR experimental values found in Fe/ZnSe/Fe junctions could be related to the nature of the interface Fe-Se bonding. Also in the case of Fe/ZnSe/Fe heterojunctions, interface engineering is probably the good approach to improve TMR.

Acknowledgments. We are thankful to Piero Torelli for usefull discussions, to Frank Vidal for the carefull reading of the manuscript. We thank also Giorgio Rossi, Jun Fujii and all the APE –ELETTRA staff for their help during experiments and for usefull discussions.

*also Dip. Fisica, Univ. Modena e Reggio Emilia, Via. A. Campi 213/A, I-41100 Modena, Italy

**Figure captions :**

Figure 1 (color online): Photoemission spectra of the Se *3d* - Fe *3p* and Zn *3d* core region recorded as a function of Fe coverage onto ZnSe(001)-C(2x2): Zn surface. Reconstructed C(2x2):Zn-ZnSe surface (in bottom) and only representative Fe coverage are shown. The topmost curve is the spectrum after removal of few atomic layers by short sputtering from the Fe surface of a 28-ML-thick Fe layer. The raw photoemission data and the deconvoluted components are shown by circles and solid lines respectively. The assignments of each deconvoluted component are presented and discussed in the text.

Figure 2 (color online): Attenuation curves of the total emission for Se *3d*, Zn 3d (filled circles), of the bulk component (open squares) and the surface component (diamond). The Fe-induced components are labelled as in Fig. 1 : Se-Fe (triangles up) and Se"floating" (triangles down) for Se 3d spectra and $Zn^0$ released (triangles up) for Zn *3d*. The Fe *3p* core-level (asymmetric line-shape) is developed at lower kinetic energy than Se *3d*. The solid line is a guide to the eyes. The attenuation of the bulk substrate core-levels expected for ideal Fe layer-by-layer (without substrate dissolution and interdiffusion) is shown for comparaison as the dashed line (in insert for Fe *3p*), calculated for an electron mean free path of 5.8 and 6.2 Å, for Se *3d* (same for Fe *3p*) and Zn *3d* emission, respectively.

Figure 3 (color online): X-ray photoelectron diffraction curves from 12 ML-Fe film on ZnSe(001) susbtrate. The intensity of Fe *3p* and Zn *3d* component of $Zn^0$ released has been recorded by scanning the polar angle with a fixed azimuth in the [100] substrate direction. The insert shows the Zn *3d* emission for clean ZnSe(001) surface and for 6 ML and 12 ML-thick Fe film.

Figure 4 (color online): Schematic picture of the atomic environment close to the Fe/ZnSe interface. Open circles, gray filled circles and black filled small circles denote Zn, Se and Fe atoms, respectively. The Fe atoms positionned near the zinc-blende sites in the bonding and antibonding positions are represented by open and filled small circles, respectively.

Figure 5 : Valence-band photoemission spectra for Fe/ZnSe (001) interface as function of Fe coverage at the photons energies of, a) 60 eV and b) 170 eV. The assignments of each peak is discussed in the text. C) $P_1/P_0$ and $P_1/P_2$ intensity ratios are plotted as a function of the Fe coverage after clean ZnSe spectra substraction.

**References**

[1] S.A. Wolf, D.D Awschalon, R.A. Burman, J.M. Daughton, S. von-Molnar, M.L. Roukes, A.Y. Chtchelkanova, and D. M. Treger, Science **294**, 1488 (2001).

[2] G.A. Prinz, Science **250**, 1092 (1990).

[3] J. M. MacLaren, X.-G. Zhang, W.H. Butler, Xindong Wang, Phys. Rev. B **59**, 5470 (1999).

[4] W. H. Butler, X.-G. Zhang, Xindong Wang, Jan van Ek, J. M. MacLaren, J. Appl. Phys. **81**, 5518 (1997).

[5] P. Mavropoulos, O. Wunnicke and P. Dederichs, Phys. Rev. B **66**, 024416 (2002).

[6] B.T. Jonker and G.A. Prinz, J. Appl. Phys. **69**, 2938 (1991); B.T. Jonker, G.A. Prinz, Y.U. Iderza, J. Vac. Sci. technol. B **9**, 2437 (1991).

[7] Y.B. Xu, E.T.M. Kernohan, D.J. Freeland, A. Ercole, M. Tselepi, J.A.C. Bland, Phys. Rev. B **58**, 890 (1998); K. H. Ploog, J. Appl. Phys. **91**, 7256 (2002).

[8] E. Reiger, E. Reinwald, G. Garreau, M. Ernst, M. Zölfl, F. Bensch, S. Bauer, H. Preis, G. Bayreuther, J. Appl. Phys. **87**, 5923 (2000).

[9] M. Marangolo, F. Gustavsson, M. Eddrief, P. Sainctavit, V.H. Etgens, V. Cros, F. Pétroff, J.M. George, P. Bencok, and N.B. Brookes, Phys. Rev. Lett. **88**, 217202 (2002).

[10] C. Lallaizon, B. Lépine, S. Abadou, S. Schussler, A. Quémerai, A. Guivarc'h, G. Jézéquel, S. Députier, R. Guérin, Appl. Surf. Sci. **123-124**, 319 (1998).

[11] V. H. Etgens, B. Capelle, L. Carbonell, M. Eddrief Appl. Phys. Lett. **75**, 2108 (1999).

[12] Xin Jiang, Alex F Panchula, Stuart S.P. Parkin, Appl. Phys. Lett. **83**, 5244 (2003).

[13] J. Varalda, A. J. A. de Oliveira, D. H. Mosca, J.-M. George, M. Eddrief, M. Marangolo, V. H. Etgens Phys. Rev. B **72**, 081302(R) (2005) ; D.H. Mosca, J.M. George, J.L. Maurice, A. Fert, M. Eddrief, V.H. Etgens, J. Magn. Magn. Mater. **226-230**, 917 (2001).

[14] F. Gustavsson, J.M. George, V.H. Etgens, M. Eddrief, Phys. Rev. B. **64**, 184422 (2001).

[15] M. Eddrief, M. Marangolo, S. Corlevi, G. M. Guichar, V.H. Etgens, R. Mattana, D.H. Mosca, F. Sirotti, Appl. Phys. Lett. **81**, 4553 (2002).

[16] E. Kisker, K. Schroder, M. Campagna, W. Gudat, Phys. Rev. B **31**, 329 (1985), Phys. Rev. Lett. **52**, 2285 (1984).

[17] M. Marangolo, F. Gustavsson, G. M. Guichar, M. Eddrief, J. Varalda, V.H. Etgens, M.Rivoire, F. Gendron, H. Magnan, D. H. Mosca, J.-M. George, Phys. Rev. B. **70**, 134404 (2004).

[18] C. Bourgognon, S. Tatarenko, J. Cibert, L. Carbonell, V. H. Etgens, M. Eddrief,

B. Gilles, A. Marty, Y. Samson Appl. Phys. Lett. **76**, 1455 (2000).

[19] R. Bertacco, M. Riva, M. Cantoni, F. Ciccacci, M. Portalupi, A. Brambilla, L. Duò, P. Vavassori, F. Gustavsson, J.-M. George, M. Marangolo, M. Eddrief, V.H. Etgens, Phys. Rev. B **69**, 54421 (2004).

[20] W. Chen, A. Kahn, P. Soukiassian, P.S. Mangat, J. Gaines, C. Ponzoni, D. Olego, Phys. Rev. B **49**, R10790 (1994).

[21] Y. Chen, F. J. García de Abajo, A. Chassé, R. X. Ynzunza, A. P. Kaduwela, M. A. Van Hove, C. S. Fadley, Phys. Rev. B **58**, 13121 (1998).

[22] A. Continenza, S. Massidda A. J. Freeman J. Magn. Magn. Mater. **78**, 195 (1989).

[23] Steven C. Erwin, Sun-Hoon Lee, Matthias Scheffler, Phys. Rev. B **65**, 205422 (2002); S. Mirbt, B. Sanyal, C. Isheden, B. Johansson, Phys. Rev. B **67**, 155421 (2003).

[24] B. Sanyal, S. Mirbt, Phys. Rev. B **65**, 144435 (2002).

[25] J. Yeh and I. Lindau, At. Data Nucl.Data Tables **32**, 1 (1985).

[26] J.A. Stroscio, D. T. Pierce, A. Davies, R. J. Celotta, Phys. Rev. Lett. **75**, 2960 (1995);P. D. Johnson, Y. Chang, N. B. Brooks, M. Weinert, J. Phys.: Condens. Matter **10**, 95 (1998).

[27] B.T. Jonker, K. H. Walker, G. A. Prinz, and C. Carbone, Phys. Lett. **57**, 142 (1986); N. B Brookes, A. Clarke, P. D. Johnson, M. Weinert, Phys. Rev. B **41**, R2643 (1990); C. Carbone, R. Rochow, L. Braicovich, R. Jungblut, T. Kachel, D. Tillmann, E. Kisker, Phys. Rev. B **41**, R3866 (1990); E. Vescovo, O. Rader, C. Carbone, Phys. Rev. B **47**, R13051 (1993); M. Sawada, A. Kimura, A. Kakizaki, Solid State Commun. **109**, 129 (1999).

[28] F. Sirotti, G. Rossi, Phys. Rev. B **49**, 15682 (1994).

[29] E. Vescovo, C. Carbone, O. Rader, Solid State Commun. **94**, 751(1995).

[30] G. Chiaia, S. De Rossi, L. Mazzolari, F. Ciccacci, Phys. Rev. B **48**, 11298 (1993); F. Ciccacci, G. Chiaia, S. De Rossi, Solid State Commun. **88**, 827(1993).

[31] A. Continenza, S. Massidda, A. J. Freeman, Phys. Rev. B **42**, 2904 (1990).

[32] E. de Jonge, P. K. de Boer, R. A. de Groot, Phys. Rev. B **60**, 5529 (1999).

[33] X.-G. Zhang, W. H. Butler, Amrit Bandyopadhyay, Phys. Rev. B **68**, 92402 (2003); E.Y. Tsymbal, K. D. Belashchenko, J. Appl. Phys. **97**, 10C910 (2005).

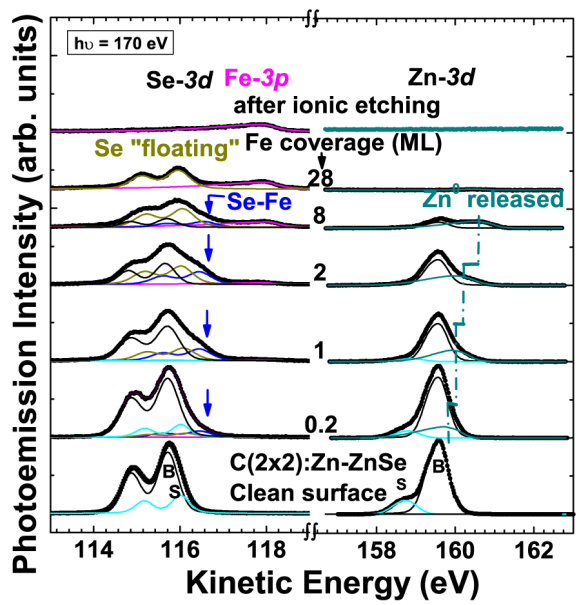

Figure 1
Eddrief et al.

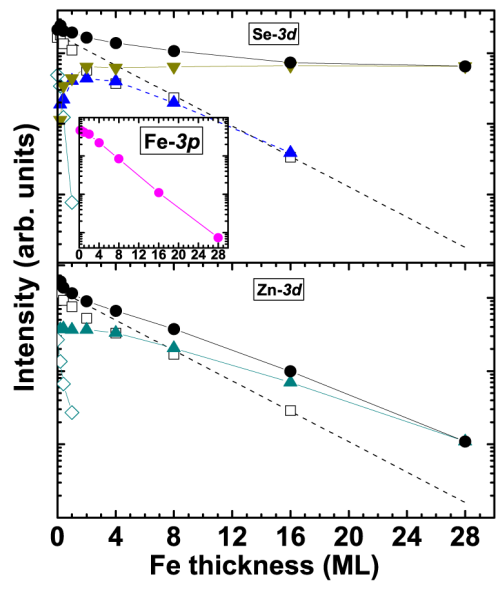

Figure 2
Eddrief et al.

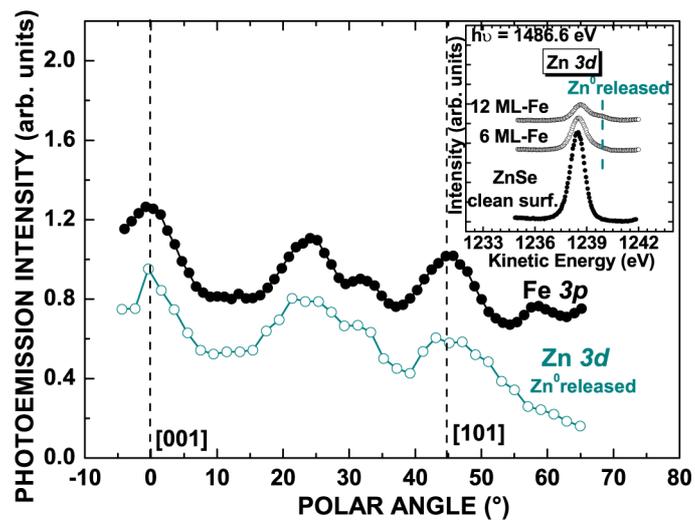

Figure 3
Eddrief et al.

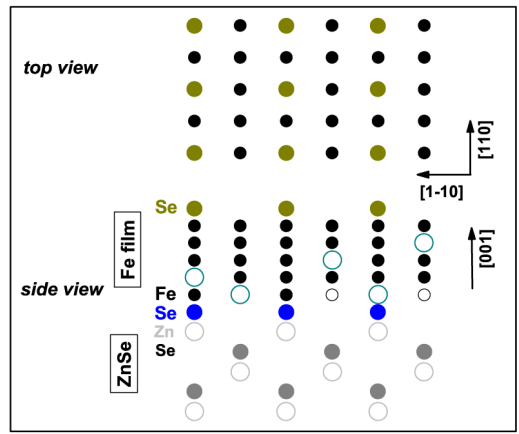

Figure 4
Eddrief et al.

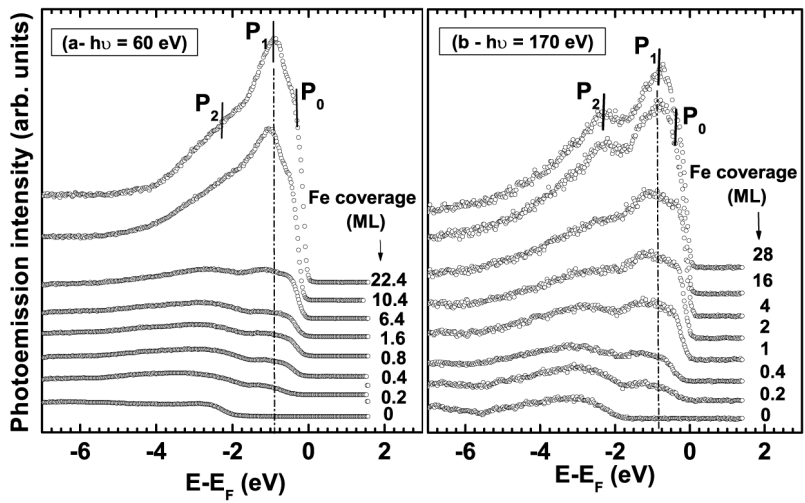
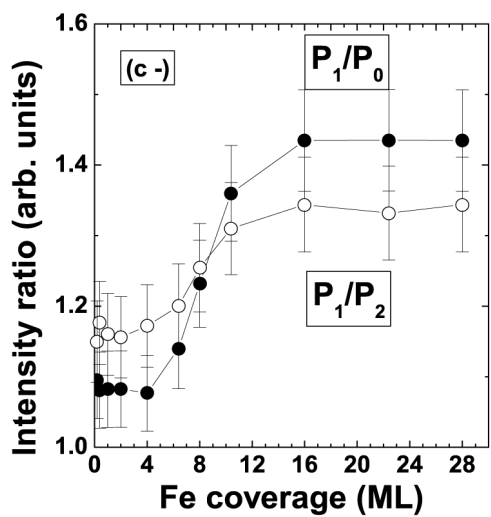